\documentstyle[epsf,prb,aps,twocolumn,graphics,floats]{revtex}

\newcommand{\be}{\begin{equation}}
\newcommand{\ee}{\end{equation}}
\newcommand{\ea}{\end{array}}

\newcommand{\bac}{\begin{array}{c}}
\newcommand{\bal}{\begin{array}{l}}
\newcommand{\baR}{\begin{array}{r}}
\newcommand{\bacc}{\begin{array}{cc}}
\newcommand{\ball}{\begin{array}{ll}}
\newcommand{\balr}{\begin{array}{lr}}
\newcommand{\barl}{\begin{array}{rl}}
\newcommand{\baccc}{\begin{array}{ccc}}
\newcommand{\barcl}{\begin{array}{rcl}}
\newcommand{\balcl}{\begin{array}{lcl}}
\newcommand{\barcll}{\begin{array}{rcll}}
\newcommand{\barll}{\begin{array}{rll}}
\newcommand{\barrclcl}{\begin{array}{rrclcl}}
\newcommand{\bacl}{\begin{array}{cl}}
\newcommand{\bacll}{\begin{array}{cll}}
\newcommand{\eac}{\end{array}}
\newcommand{\ber}{\begin{eqnarray}}
\newcommand{\eer}{\end{eqnarray}}

\newcommand{\ca}{Ca$_{14}$MnBi$_{11}$}
\newcommand{\ba}{Ba$_{14}$MnBi$_{11}$}

\begin{document}
\draft
\flushbottom
\twocolumn[\hsize\textwidth\columnwidth\hsize\csname @twocolumnfalse\endcsname

\title{Bonding, Moment Formation, and Magnetic Interactions \\
                    in \ca~and \ba} 

\author{D. S\'anchez-Portal$^{1,2}$, Richard M. Martin,$^1$ 
         S. M. Kauzlarich,$^3$ and W. E. Pickett$^4$}

\address{$^1$Department of Physics and Materials Research Laboratory, 
University of Illinois, Urbana, Illinois 61801}
\address{$^2$Departamento de F\'{\i}sica de Materiales and DIPC, 
Facultad de Qu\'{\i}micas, UPV/EHU, Apdo. 1072, 
E-20080 San Sebasti\'an, Spain}
\address{$^3$Department of Chemistry, 
University of California, Davis CA 95616}
\address{$^4$Department of Physics, 
University of California, Davis CA 95616}

\date{\today}
\maketitle


\begin{abstract}
The ``14-1-11'' phase compounds based on magnetic Mn ions and typified by
\ca~and \ba~show unusual magnetic behavior, but the large number (104)
of atoms in the primitive
cell has precluded any previous full electronic structure study.
Using an efficient, local orbital based method within the local
spin density approximation to study the electronic structure, we 
find a gap between a bonding valence band complex and an antibonding
conduction band continuum.  The bonding bands lack one electron
per formula unit of being filled,
making them low carrier density $p$-type metals. 
The hole resides in the MnBi$_4$ tetrahedral unit
and partially compensates the high spin $d^5$ Mn moment, leaving a
net spin near 4 $\mu_B$ that is consistent with experiment.
These manganites are composed of 
two disjoint but interpenetrating `jungle gym' networks 
of spin $\frac{4}{2}$ MnBi$_4$$^{9-}$ units 
with ferromagnetic interactions within the same network,
and weaker couplings between the 
networks whose sign and magnitude is sensitive to 
materials parameters.
\ca~is calculated to be ferromagnetic as observed, while for \ba~(which is
antiferromagnetic) the ferro- and antiferromagnetic 
states are calculated
to be essentially degenerate. 
The band structure of the 
ferromagnetic states is very close to half metallic.
\end{abstract}
\vskip 0.5cm

]

\section{Introduction}

Recently many new magnetic phenomena have been discovered, 
such as colossal magnetoresistance\cite{book} (CMR),
a new type of 
heavy fermion system\cite{livo} (LiV$_2$O$_4$), 
and spin Peierls ground states in magnetic 
insulators. Spin glass behavior
has been found and studied intensively
in magnetic insulators without structural
disorder\cite{y2mo2o7}.
A related aspect of complexity in crystalline magnets arises when magnetic 
ions are distributed regularly but interionic distances are large, 
$\sim$1~nm.
One example 
is the class of metallic rare earth (RE) 
silicides
(RE)$_3$Pd$_{20}$Si$_6$, and the corresponding germanides, 
which present multiple ordering transitions with temperatures\cite{silicides}
in the range of a few K.
Examples of dilute magnetic insulators are 
Na$_3$M$_2$Li$_3$F$_{12}$\cite{fluorides}
and the rare earth phosphomolybdates\cite{moly} 
(RE)PO$_4$(MoO$_3$)$_{12} \times$
30H$_2$O, which order below 1 K, sometimes much below.  A more well known
example is the heavy fermion superconductor UBe$_{13}$, which superconducts
at 0.9 K and is near an antiferromagnetic instability\cite{UBe13}.
These intriguing materials, with their complex exchange interactions 
transmitted through intermediate bonds,
are examples of the continuing, perhaps even
accelerating, growth and study of more complex compounds.  The complexity
may be structural in origin (many atomic sites of low symmetry in the
unit cell), or it may be rooted in complex magnetic and electronic interactions
(many different exchange couplings, perhaps competing and frustrating).

In this paper we present a first-principles density-functional 
study of the electronic 
properties of two representative compounds of the so-called 
``14-1-11'' phases, which are a clear example of materials showing 
simultaneous complex atomic and magnetic structures.
The 14-1-11 phases~\cite{kauz} typified by \ca, where Ca may be 
substituted by Sr or Ba
and Bi can be substituted by Sb or As,\cite{rehr} are a 
rather difficult case of a magnetic system with a
complex crystal structure.  The structure will be described in detail below,
but with its four formula units (104 atoms) and nine inequivalent sites,
and magnetic ordering at 15-70~K indicating exchange coupling of the
order of 10 meV,
this class presents a strenuous test for state of the art electronic structure 
methods.  Understanding the existence of such a structure is itself an
interesting topic in solid state chemistry. It seems to be consistent with
the simplified picture provided by the 
Zintl-Klemm-Bussmann\cite{ZKB} concept, which we will refer to as the
Zintl concept.  This picture represents a generalization of the octet rule
for binary semiconductors and insulators, and invokes charge balance between
(nominally closed shell) structural units which themselves may be 
covalently or ionically bonded complexes as well as simple ions.
However, both of the compounds that we address in detail in this paper are
metallic~\cite{kuro,heatcapacity} in their magnetically
ordered phases, whereas the Zintl arguments 
that have been applied to these compounds
would be more
appropriate if they were semiconducting.
In fact, the magnetic ordering and
the metallicity seem to be correlated in these
materials, which would be broadly consistent with a
Ruderman-Kittel-Kasuya-Yosida (RKKY)
type of model~\cite{RKKY} for the origin of the magnetic
interactions.

The two materials chosen for the 
present study are \ca, which is 
ferromagnetic (FM) with Curie temperature T$_C$=55~K, and antiferromagnetic
(AFM) \ba, with N\'eel temperature T$_N$=15~K.  The observed
magnetic moment of Mn and the ordering temperatures (several tens of K)
give rise to several fundamental questions: \\
(1) what is the charge state of Mn, and how does it relate to structural
stability and conduction behavior?\\
(2) given that the Mn ions are
magnetic, why is the ordering temperature of the order of 50~K when the 
distance between Mn ions is at least 10~\AA?  \\
(3) what is the bonding path that provides the 
magnetic coupling? \\
(4) how is the magnetic order coupled to the carriers in these compounds,
some members of which show colossal magnetoresistance 
near T$_C$? \\  

The only previous theoretical work on the electronic structure 
of this class of compounds was
done by Gallup, Fong, and Kauzlarich,~\cite{gallup} who considered a 
single-formula-unit simplification of semiconducting Ca$_{14}$GaAs$_{11}$.
They concluded that the bonding in this compound is consistent
with the Zintl concept of valence counting using covalently bonded
subunits (discussed below).  The greatest difference between
this compound and those we study in this paper 
is the substitution of the $sp$
metal atom Ga in the site of inversion symmetry
with the transition metal atom Mn. Ga
is trivalent, whereas the metallicity of the Mn-based compounds
indicates a different valence for Mn.

We conclude, in fact, 
that Mn is in a divalent state, with five $3d$ 
electrons localized and magnetic.
This difference, trivalent Ga versus divalent Mn, leaves one
unoccupied bonding orbital in the valence bands, giving metallic
behavior as observed. The FM metal \ca~ is close to a half metallic
filling\cite{physicstoday} of the bands. 
Our results suggest that adding one additional electron
per formula unit to the itinerant valence bands should lead to a 
semiconducting compound.  Adding less than one more carrier could
produce a half metallic FM situation. 
The most direct way of doing so would be the replacement of a fraction of
alkaline earth atoms with a trivalent atom, {\it viz.} 
Ca$\rightarrow$Y.  A less likely possibility would be Mn$\rightarrow$Fe,
if Fe would assume a trivalent, high spin $d^5$ configuration.
The alternative is that 
Fe would assume a $d^6$ configuration,
and therefore also be divalent, and the system would remain metallic.

Current interpretation assigns
a 3+ valence to the Mn atoms in these 
compounds~\cite{reviewZintl,prbFisher}.
This disagrees
with our findings as stated above, but it 
seems a very reasonable assumption considering
that the measured magnetic moment 
is very close to 4~$\mu_B$/Mn for this
class of materials.
However, Mn$^{3+}$ also seems to 
imply a semiconducting character ({\it viz.} 
Ca$_{14}$GaAs$_{11}$), in 
contradiction to the experimental evidence.
While the valence (or charge state) of an ion in a solid is a very 
useful concept, it does not necessarily represent an actual ionic charge;
indeed, it is widely recognized that `ionic charge' is an ill defined
concept, particularly so for metals and narrow 
gap semiconductors.  Nevertheless, the `charge state' often continues to
be meaningful, and
the charge state of the Mn atom in these materials has been 
deduced from measurements of the high temperature magnetic suceptibility,
which is indicative of moments of $\sim$4~$\mu_B$. 
This moment was assigned to Mn, arriving at a $d^4$ (trivalent) assignment. 
We find however that the density 
of states (DOS) contains a peak, well below the Fermi energy, 
originating from five bands per formula unit ( i.e. per Mn atom)
which derive from Mn 3$d$ states.
This situation allows
us to identify the Mn ion as 
having a $d^5$ 
configuration, consistent only with a divalent charge state.
Nevertheless, the moment that we obtain
is consistent with experiment, being closer to 4~$\mu_B$ than 
to the 5~$\mu_B$ expected from a simple $d^5$ configuration. 

These two apparently contradictory observations 
are reconciled by the near-half-metallic band structure: 
the holes (one per Mn atom)
left in the valence band reside in the Bi$_4$ tetrahedron that
encapsulates the Mn ion.
The effective
$\sim$4~$\mu_B$ magnetic moment is still relatively well localized.
As we will see below in more detail, both the experimental 
moment of $\sim$4~$\mu_B$ and our observation of an almost half metallic
band structure
can be already anticipated from 
the electronic structure
of the isolated MnBi$_4^{-n}$ tetrahedron,
using the charge derived from applying  
the valence counting rules (n=9).

The paper is organized as follows.  Sections~II and III describe the
crystal structure and method of calculation, respectively.  Magnetic
energies and their interpretation in terms of exchange couplings are
presented in Sec.~IV, and Sec.~V presents an analysis of the 
magnetization and its 
distribution along the unit cell.
The density of states of
crystalline \ca~ and \ba~ are analyzed in Sec.~VI, followed by a
discussion of charge transfers
and their
relation with the formal
valences in Sec.~VII.  Sec.~VIII is
devoted to examine the electronic states near
Fermi level along with their relation with the anysotropic nature of the
magnetic couplings. In Sec.~IX the main results for the solid
are reinterpreted as simple consequences 
of the electronic structure 
of the isolated MnBi$_4$ charged clusters, 
and a summary is presented in Sec.~X.

\section{Crystal Structure and Its Implications}
\label{structure}

\begin{figure}[tbp]
\epsfxsize=6.0cm\centerline{\epsffile{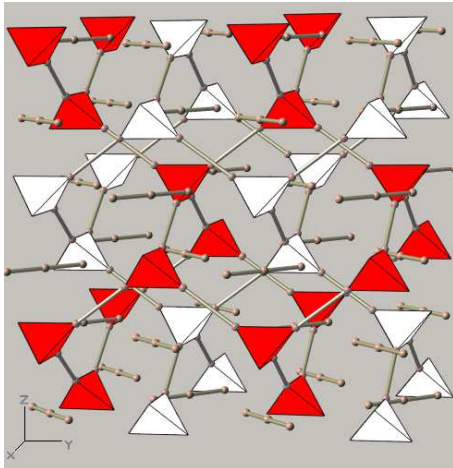}}
\caption{One view of the crystal structure of \ca, emphasizing the two
interpenetrating sublattices of MnBI$_4$ tetrahedra, one shown as white and
the other one darker.  The Bi$_3$ ``sticks'' are also shown; the Ca atoms
and the isolated Bi atoms are not shown, for clarity. 
\label{Fig1}}
\end{figure}

The structure of the alkaline earth metal pnictide compounds 
Ae$_{14}$MPn$_{11}$, which we will call the AeMPn structure, 
has been experimentally 
determined, and described in detail 
by Kurotomo, Kauzlarich, and Webb.\cite{kuro} In the present
calculations we have used the
coordinates given by these authors.
Ae$_{14}$MPn$_{11}$ compounds
have a body center tetragonal (space group I4$_{1}$/acd)
unit cell with 4 formula units (104 atoms), 
a=17.002~\AA, c=22.422~\AA\ for \ca, 
and a=18.633~\AA, c=24.340~\AA\ for \ba.\cite{kuro}
The structure can
be viewed as consisting of: interstitial alkaline earth atoms, 
isolated Bi (Bi3\cite{nomenclature}) atoms, 
distorted MnBi$_4$ (Bi2) tetrahedra, and Bi$_3$ (Bi1-Bi4-Bi1) linear units. 
The MnBi$_4$ tetrahedra are translated by $1\over2$ along the $c$ axis
alternating with the Bi$_3$ anions which are rotated by 90$^{\rm o}$
with respect to each other, as shown in {Fig.~\ref{Fig1}}. The
isolated Bi atoms are situated between the Bi$_3$ and MnBi$_4$ groups,
along a screw axis which coincides with the $c$ axis.
All the Mn atoms in the unit cell are symmetry equivalent. 
The alkaline earth cations occupy four inequivalent sites, but the
distinctions will not concern us. 

The MnBi$_4$ tetrahedra are slightly flattened within 
the $a$-$b$ plane, with the distortion increasing by $\sim$1$^{\rm o}$ when 
Ca ion is substituted by Ba (the inequivalent angles are
118.0$^{\rm o}$ and 105.4$^{\rm o}$ for the Ca compound). 
All bond lengths in the system
also increase with this substitution, for which $a$ and $c$ increase
by 9-10\%, with a corresponding volume change of nearly 30\%.  The Mn-Bi bond
distance, for example, increases from 2.814~\AA\ (Ca) to 
2.935~\AA\ (Ba). The Bi-Bi
bond length in the Bi$_3$ units also increases from 3.335~\AA\ (Ca) to
3.498~\AA\ (Ba)~\cite{kuro}, indicating that both this unit and the
MnBi$_4$ tetrahedra are substantially environment dependent.       

There are four cations located close to the Bi2 atoms in the tetrahedron,
with Bi-Ca distances in the range 3.2-3.3\AA, and  3.5-3.6~\AA\ for 
Bi-Ba separations~\cite{kuro}. 
The Bi4 central atom in the Bi$_3$ unit is surrounded by
four cations, while the terminal Bi1 atoms are coordinated with eight cations 
(Ca-Bi bonds of 3.2-3.4~\AA, Ba-Bi distances of 3.5-3.75~\AA)\cite{kuro}.

As pointed out in Ref.~\onlinecite{reviewZintl}, 
and will prove
to be important to understand the magnetic couplings,
this crystalline structure can also be regarded as 
two interpenetrating networks
formed by MnBi$_4$ tetrahedra.
Each tetrahedron only belongs to one of these networks, and 
is linked to the four nearest 
MnBi$_4$ groups in the same network along the 
tetrahedral directions, i.e. through a Bi-Bi bond.
This loose bond, with the Bi atoms
4.5-5~\AA\ apart,
is mediated by three cations
(atoms A1, A1$^{\prime}$, and A3 in Ref.~\onlinecite{kuro}),
which are common
nearest neighbors to both Bi (Bi2) atoms.
Fig.~\ref{Fig2} shows a schematic representation 
of this connectivity on the (100) plane.

In contrast, the identification of the
interaction pathway between MnBi$_4$ tetrahedra
belonging to different networks is not so clear.
Different
paths can be envisioned, all involving several cations and  
at least one of the isolated Bi atoms (Bi3).

For compounds such as Ae$_{14}$GaPn$_{11}$, 
formal valence arguments and accumulated experience indicate 
that the electronic structure can be rationalized in terms
of a model where each 
alkaline earth atom cedes two electrons (14A$^{2+}$),
three of these valence electrons are collected by each isolated
Pn atom (4Pn$^{3-}$), 
seven electrons can be transferred to the 
Pn$_3$ units leaving unoccupied the antibonding $\sigma$ molecular 
state~\cite{gallup}, and the remaining elecrons go to 
the tetrahedron centered on the Ga atom which becomes
GaPn$_4^{9-}$. Although 
these formal charges are quite large and should not be taken literally, 
such valence counting has proven 
to provide a very good description of the electronic structure 
of Ca$_{14}$GaAs$_{11}$~\cite{gallup}. 

The metallic or semiconducting
character of the Ae$_{14}$MnBi$_{11}$ materials therefore seems to depend in
an essential way on 
the valence of the Mn atom.
A trivalent Mn atom is isovalent with Ga and implies
a semiconducting compound (which the Ae$_{14}$MnBi$_{11}$ 
compounds are not), 
while a divalent Mn ion will imply a metallic compound (as observed) but
appears to violate formal charge neutrality.
One of the main objectives
of this paper it to resolve this conundrum.

\section{Method of Calculation}
The calculations have been performed with the program
SIESTA,\cite{SIESTA1,SIESTA2,SIESTA3} an optimized
code which allows  standard density-functional~\cite{DFT}
calculations
on systems with hundreds of atoms. 
This method has been successfully applied to the study of 
the electronic and structural properties
of many different materials\cite{reviewPablo}, including magnetic
clusters\cite{clusters}.
The computational
cost and memory requirements
to build up the Hamiltonian matrix scale linearly with the size
of the system, {\it i.e.} is of
order N ($O(N)$) where N is the number of atoms,\cite{SIESTA1,SIESTA2} 
while for the 
evaluation of the density and energy one can choose between standard
methods, or make use of the recently
developed $O(N)$ techniques.~\cite{O(N)} In this work we
have used a standard diagonalization of the Hamiltonian because we are
interested in the band structure and characteristics of specific states.

\begin{table}[tbp]
\caption[]{ Core radii (in a.u.)
used for the generation of the pseudopotentials}
\begin{tabular}{ccccc}
        & Mn & Bi   & Ca & Ba \\
r$_s$ & 2.00 & 2.20 & 3.20 & 3.50  \\
r$_p$ & 2.20 & 2.90 & 3.30 & 4.00 \\
r$_d$ & 1.90 & 2.90 & 3.00 & 3.50
\end{tabular}
\label{pseudo}
\end{table}

The basis set is a
linear combination of pseudoatomic orbitals.~\cite{SIESTA3,Sankey,PAO}
In the present calculations we have used a
double-$\zeta$
polarized~\cite{SIESTA3,javier,double,polarization} (DZP) basis set for all
the atoms. A shell of $d$ orbitals was also included for
the Ba and Ca species.
This amounts to 15 orbitals (2 $s$, 10 $d$, and a
polarization (P) shell with 3 $p$ orbitals) for Mn, 13 orbitals
( 2 $s$, 6 $p$, and 5 P $d$) for Bi, and 10 orbitals ( 2 $s$, 5 $d$,
and 3 P $p$) in the case of Ca and Ba, making 1192 orbitals in total.

The core electrons are replaced by norm-conserving
pseudopotentials\cite{Troullier-Martins} generated from the
atomic configurations [Ar]3$d^5$4$s^2$ for Mn,
[Hg]6$p^{1.75}$6$d^{0.25}$5$f^{0.25}$ for Bi~\cite{bachelet} and,
[Ar]4$s^1$ and [Xe]6$s^1$ for Ca and Ba respectively.
The core radii used in the generation of the pseudopotentials 
can be found in Table~\ref{pseudo}.
We apply the pseudopotentials using the fully separable formulation
of Kleinman-Bylander.~\cite{Kleinman-Bylander}

The calculations have been carried out in the local spin density
approximation (LSDA)~\cite{cepald},
and a partial-core correction for the non-linear
exchange-correlation~\cite{NLCC} has
been included for all the species. This correction is specially
important to get reliable moments in magnetic atoms, but
some care has to be taken in choosing the pseudocore radius.
We found that a pseudocore radius of 0.70 a.u. 
leads to very good
results in comparison to all-electron calculations.~\cite{Ruben}

\section{Results for Total Energies and Magnetic Couplings}

\begin{figure}[tbp]
\epsfxsize=6.0cm\centerline{\epsffile{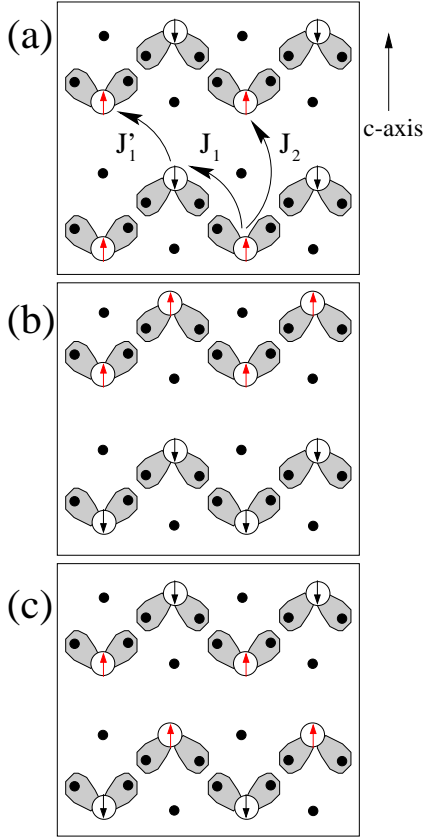}}
\vskip 3mm
\caption{Illustrations of the three inequivalent types of AFM order
possible without enlarging the unit cell.  A (100) plane is shown.
The open circles with arrows indicate Mn with its spin direction, the
filled dots indicate Bi sites, and the grey lobes indicate bonding
patterns schematically.  Some of the exchange constants are shown.
}
\label{Fig2}
\end{figure}

In this section we report on the relative 
stability of FM and AFM alignments
of the Mn moments and  
the magnetic couplings 
deduced from the corresponding 
values of the total energy
in both compounds.
Since the calculations are already very large with the primitive
crystallographic cell, we have not considered any magnetic alignments
that would enlarge the cell.  Even so, with four magnetic Mn atoms
in the cell there are three distinct types of AFM alignments, shown
in {Fig.~\ref{Fig2}} in the (100) plane.  The configurations
are better explained in terms of two distinct 
...-Mn-Bi-Bi-Mn-... `chains' in this plane:
(i) in-phase AFM chains (Fig.~\ref{Fig2}(a)), 
(ii) antialigned FM chains (Fig.~\ref{Fig2}(b)), 
and (iii) 
out-of-phase AFM chains (Fig.~\ref{Fig2}(c)). 
Similar chains run out of this plane
and ultimately the various bonding `chains' form two interpenetrating
three dimensional
networks, as described in Section~\ref{structure}.
Only four of the eight nearest 
Mn neighbors of a given Mn atom
belong to the same network as the central one, and only 
in the configuration (i)
(in-phase AFM chains) the eight neighbors
are antiferromagnetically aligned. 

From the figure it is clear that there are two distinct nearest neighbor
interactions, labelled J$_1$ and J$_1^{\prime}$.  The second 
neighbor interaction J$_2$ is also shown.  We will obtain an estimate of
these couplings assuming an effective spin coupling
\begin{equation}
\label{heff}
H_{spin} = -\sum_t \sum_{<ij>} J^{(t)}_{ij} S^z_i S^z_j,
\end{equation}
where $S^z_i = \pm 1$ represents a normalized spin variable.  The index $t$
labels the types of couplings, and $<ij>$ denotes spin pairs of
that type.  Note that in
some treatments, our J might correspond to JS$^2$ where S is the spin of
the magnetic atom.

For both compounds we have calculated the energies for 
FM order and for the three
AFM configurations decribed above. 
Of the AFM alignments, the antialignment
of FM chains, as shown in Fig.~\ref{Fig2}(b), 
is always the one favored.  
Therefore, the minimum energy AFM configuration
occurs when one of the Mn networks is entirely spin
up, the other entirely spin down.  

Since the energy differences are quite small
(as expected), it was necessary to study the convergence of the energies
with respect to both the number of points (cutoff~\cite{cutoffmeaning})
in the real-space
grid,
and the k-point mesh used to sample the Brillouin
zone.  Cutoffs
up to 216~Ry for \ca~and 147~Ry for \ba~were
used, with up to 12 k-points in the irreducible Brillouin zone.
The results shown in the following have been obtained
with 3 inequivalent k-points, and a cutoff of 150~Ry and 147~Ry
for the Ca and Ba compounds, respectively.
With these parameters the total energies
are converged, for a given basis set,
to 10~meV/Mn, and 5~meV/Mn for the
energy differences between different configurations, which show 
a faster convergence.

\begin{table}[tbp]
\caption[]{Upper: Energies of the three antiferromagnetic alignments
pictured in {Fig.~\ref{Fig2}} relative to the ferromagnetic
alignment.  The type of AFM order in \ba~(type (a), (b), (c), or
other, is not established experimentally).  Numerical uncertainties
are expected to be $\sim \pm$ 5 meV.
Lower: Exchange couplings for a short
range Ising model determined
from the total energy differences.
The couplings correspond to those pictured in {Fig.~\ref{Fig2}}.
J$_{2+3}$ are the value of the couplings if
second and third neighbors would enter in the energy expression
via a single exchange constant (see the text).
}
\begin{tabular}{ccc}
 Energy (meV)      & \ca             &  \ba  \\
\tableline
  AFM(a) - FM      &   59            &   12  \\
  AFM(b) - FM      &   25            &    4  \\
  AFM(c) - FM      &   52            &   13  \\
\tableline
 Experiment        &    FM           &  AFM  \\
\tableline
\tableline
Exchange Constants (K)   & \ca             &  \ba  \\
\tableline
  $J_1$            &   63 $\pm$~5     &  15 $\pm$~5   \\
  $J_1^{\prime}$   &   23 $\pm$~5     &  ~2 $\pm$~5   \\
  $J_2$            &   27 $\pm$10    &  ~7 $\pm$10  \\
  $J_{2+3}$        &   ~9 $\pm$~3     &  ~2 $\pm$~3    \\
\tableline
T$_C$, mean field  &   130$\pm$20 K  &  28$\pm$20 K \\
T$_C$, experiment  &    T$_C$=55 K   &  T$_N$=15 K

\end{tabular}
\label{table1}
\end{table}

The calculated total energy differences are presented in
{Table~\ref{table1}}.
For \ca~the FM state is more
stable by 25$\pm$5 meV per Mn atom than the lowest energy
AFM phase, in accord with the observation of ferromagnetism in this
compound.  In a simple nearest neighbor
Ising or Heisenberg modelling of the spin coupling 
this energy would
correspond to a FM exchange coupling of J$\sim$3~meV.  For \ba~the energy
difference is much smaller, with the FM phase 
4$\pm$5 meV/Mn more stable, {\it i.e.} degenerate to within our accuracy.
This compound is observed to be AFM, with
T$_N$=15 K, with the low ordering temperature reflecting smaller
magnetic interactions than in \ca.

{Table~\ref{table1}} also presents the exchange couplings of 
the effective spin Hamiltonian (\ref{heff}).
Each Mn atom has eight Mn first neighbors, four on the (100) and 
four on the (010) planes, 
at distances of 10.18~\AA\ for the Ca compound (10.84~\AA\ for
\ba).
Already from the energies in {Table~\ref{table1}}, it is evident that
two different first neighbor couplings need to be included. 
In fact, 
if just one exchange constant were to be used in a model with 
only first neigbor interactions, the energies of configurations
(b) and (c) in {Fig.~\ref{Fig2}} should be identical.
Physically, the need of two constants is
easily understood as a consequence of 
the presence of two Mn networks:
stronger interactions can be expected 
within the same
network (J$_1$) than between atoms in different networks (J$^{\prime}_1$). 
This is clearly confirmed in {Table~\ref{table1}}, where
J$_1$ is shown to be much lager than J$^{\prime}_1$ for both compounds.

There are two
second neighbors located at 11.21~\AA\ (12.17~\AA ) along the c axis.
The four third neighbors
can be found
12.02~\AA\ (13.18~\AA ) away,
along the $<$100$>$ and equivalent directions.
In the crystallographic cell these last atoms
are equivalent to the second neighbors yet located at
somewhat larger distances due to the tetragonal distortion
of the cell. Further neighbors are at more than 17~\AA\ for both compounds
and their interaction constants are expected to be much smaller.
With the calculated energy differences
we cannot make independent estimations
of the second (J$_2$) and third (J$_3$) nearest neighbor 
couplings. At this point, we can choose 
to restrict the interactions
up to second neighbors and obtain a value for 
J$_2$. Another possibility
is to 
consider
second and third neighbors as entering in (\ref{heff}) at the
same footing, i.e. through an {\it averaged} coupling constant
(J$_{2+3}$=(2J$_2$+4J$_3$)/6). The values for these exchange constants
are also listed in {Table~\ref{table1}}.

\section{Magnitude and Distribution of the Magnetic Moment}
\label{momentsection}

In this section we study the distribution of the magnetic moment
in the unit cell with the help of the Mulliken population 
analysis\cite{Mulliken}. 
It must be recognized that the populations
obtained using Mulliken
analysis suffer, as those obtained via
all similar techniques, from an inherent
arbitrariness: both the charge density and the total charge
are observables of the
system, but their partition into different atomic-like contributions
cannot be uniquely defined. Accordingly, Mulliken
population trends and differences are more meaningful that their
absolute values, and their use is fairly common.
We obtain in this way
information about the effective valence state and magnetic moment
of the Mn atoms, 
the polarization of the surrounding atoms, 
and the degree of localization of the total 
magnetic moment.

The magnetic moments on the Mn atoms for the FM compounds, 
as obtained from the difference between the Mulliken populations of
the majority and minority spin densities, 
are shown
in {Table~\ref{population}}. 
They are found to be nearly independent
of the type of magnetic order, and differ by 4\%:
4.45~$\mu_B$ for \ca, 4.62~$\mu_B$ for \ba.  
The part of the moment
attributable to the 3$d$ shell is 4.16 $\mu_B$ and 4.32 $\mu_B$,
respectively, with about 0.1 $\mu_B$ and 0.2 $\mu_B$ induced in the
4$s$ and 4$p$ states in each compound. 
The spin up population ({Table~\ref{population}}) of the Mn 3$d$ orbitals 
approaches 5 electrons, indicative of 
a $d^5$ configuration. Several factors can contribute 
to a reduction of the 3$d$ population, being the most important
the hybridization with the 6$p$ states
of the neighboring Bi atoms. 
The identification of the configuration of Mn 
as 3$d^5$ will be also confirmed from a detailed analysis of the electronic
band structure (see Section~\ref{MnCharacter}), providing further
support
to our picture of a divalent Mn atom.

The total moment per formula unit of the FM state is calculated 
to be (see {Table~\ref{population}}) 4.25 $\mu_B$ for \ca~
and 4.40 $\mu_B$ for \ba, which becomes 17 $\mu_B$ and 17.6 $\mu_B$
per cell, respectively. 
With all the  majority $d$ states occupied,
each Mn atom should contribute with a magnetic moment of +5~$\mu_B$ 
(+20~$\mu_B$ in the unit cell). 
However, analysis of the contributions to 
the $d$ bands indicates that this 
moment, 
although still mostly localized in Mn, 
is somewhat spread out to the neighboring Bi atoms (Bi2)
due to the considerable 
hybridization between the Mn 3$d$ states and the 6$p$ states
of Bi.
Therefore, it is
more safely pictured as associated with the five atoms in the
MnBi$_4$ tetrahedra, rather than to the Mn atoms alone.

\begin{table}[tbp]
\caption[]{
Population and moments of the Mn $d$ shell,
and magnetic moments of Mn atoms and MnBi$_4$ tetrahedra as
obtained from a Mulliken population analysis for the
FM \ca, and \ba.
$ \mu_{Ae_{14}MnBi_{11}}$ stands for the net magnetic
moment per formula unit, while 
$\mu_{eff}^{exp}$ is the effective
moment per Mn atom as obtained
from the measurements of the
high temperature magnetic suceptibility. }
\begin{tabular}{lcc}
& \ca &   \ba \\
\tableline
Q$^{\uparrow}_d$   &  4.67 &   4.74 \\
Q$^{\downarrow}_d$  &  0.51 &   0.42 \\
$\mu_d$          &  4.16 &  4.32  \\
$\mu_{Mn}$       &  4.45 &  4.62  \\
$ \mu_{MnBi_4}$  &  4.22 & 4.35   \\
$ \mu_{Ae_{14}MnBi_{11}}$&  4.25 & 4.40   \\
\tableline
$\mu^{exp}_{eff}$&  4.8  & 4.8
\end{tabular}
\label{population}
\end{table}

A reverse polarization of
2.4--3 $\mu_B$ (0.6--0.75 $\mu_B$/MnBi$_4$ unit)
must be present to recover the
calculated moment.
From the data in {Table~\ref{population}}
it is clear that this back polarization is also
concentrated in the MnBi$_4$ tetrahedra, whose magnetic moment 
almost recover the total value per formula unit. Part
is due to the occupation,  through hybridization, 
of the minority spin Mn $d$ orbitals.  The remainder
is mostly divided among the Bi2 atoms, which 
have a magnetic
moment of  $\sim-0.06~\mu_B$ per Bi atom, the largest 
in the cell after Mn.
This moment is quite modest, but it is in fact built up
from two much larger contributions that tend to cancel each other:
$\sim+0.20~\mu_B$ coming from their participation in 
the majority Mn 3$d$ bands (as previously commented), 
and a $\sim-0.26~\mu_B$ component
coming from the polarized Bi 6$p$ bands reponsible for the 
reverse polarization.
The amount of spin polarization spread around the rest 
of the cell is negligible 
($\sim0.001~\mu_B$  per Bi or Ae atom on average). 
Only the isolated Bi atoms (Bi3), 
and the alkaline earth atoms (A1 and A3)
connecting the MnBi$_4$ tetrahedra present a small polarization 
of $\sim-0.03~\mu_B$ and $\sim+0.02~\mu_B$, respectively.

An analysis of the Mulliken populations of all the 
studied AFM orders, for both Ca and Ba compounds, 
indicates a very similar picture: the calculated moment
per formula unit has to be assigned
to the MnBi$_4$ tetrahedra, being neither fully 
localized in the Mn ion nor spread over the whole unit cell. This 
moment, of $\sim$4$\mu_B$, is 
built up from two distinct contributions, a $\sim$5$\mu_B$ moment
coming from the fully occupied majority spin 3$d$ shell of Mn, 
and a 
$\sim-1$$\mu_B$ reverse polarization coming from 
the 6$p$ states of the four Bi atoms in the MnBi$_4$ complex.
The MnBi$_4$ tetrahedra are therefore the {\it magnetic units}
in these compounds.

Experimentally, an effective moment ($\mu_{eff}$)
of 4.8~$\mu_B$ has been obtained
for both compounds 
from the high temperature suceptibility~\cite{kuro}.
In fact, for all the related compounds 
Ae$_{14}$MnPn$_{11}$, 
where Bi is substituted
by Sb or As, and the cation sites are occupied by Ca, Sr or Ba atoms,
$\mu_{eff}$ is always found within the range 
4.8--5.4~$\mu_B$~\cite{rehr,kuro}. For
Yb$_{14}$MnBi$_{11}$ and Yb$_{14}$MnSb$_{11}$~\cite{reviewZintl,prbFisher}
$\mu_{eff}$ is also measured to be 4.9~$\mu_B$.
This has been interpreted, using the standard form 
for the Curie constant, $\mu^2_{eff}$=g$^2$J(J+1)$\mu_B^2$ where
J is the total effective angular moment and g is the Land\'e 
g-factor\cite{Ashcroft},
as a signature of the 
presence of localized $\sim$4~$\mu_B$ moments in all these compounds.
This seems
in good agreement with our calculated
magnetic moment per formula unit (i.e. per MnBi$_4$ tetrahedron).
However, all previous works have assigned these magnetic moments solely 
to the Mn atoms, leading to the conclusion
that Mn is in a 3+ valence state with four unpaired $d$ electrons,
which
is in contradiction by our findings.

A better quantity to compare
with our calculation of the magnetic 
moment is the experimental 
saturation moment, which is the value of the ordered moment parallel to
the applied field $<$M$_z$$>$.
The initial work on hysteresis loops for powder samples,~\cite{rehr,kuro} 
observed saturation moments of 2.5--3~$\mu_B$/Mn,
which are $\sim$30\% smaller than both the estimations based on 
the measured Curie constants, and our calculated total moments.
The polarization of 
the itinerant valence electrons was invoked in those works
to
explain this discrepancy between the measured 
saturation moments and the fitted $\mu_{eff}$.
However, more recent measurements
using single crystals~\cite{cmr-webb,prbFisher}
indicate that the saturation moment 
is close to 4~$\mu_B$/Mn, in better agreement with our calculations.

\section{Decomposition of the Electronic Spectrum}

Besides the character of the band structure near the Fermi energy,
which will be discussed in detail in a separate section,
the general features of the electronic structure and
ordering of the levels are the same for both FM and AFM orders,
and very similar for Ca and Ba compounds. 
Therefore we will concentrate here
in the FM order. 

\subsection{Total Density of States, General Structure}
\label{TotDOS}

With five valence electrons for Bi, two for Ca, and seven for Mn, and
four formula units per cell, the occupied valence bands of these
compounds must accommodate 360 electrons, or roughly 180 occupied
bands of each spin.  

\begin{figure}[tbp]
\epsfxsize=7.0cm\centerline{\epsffile{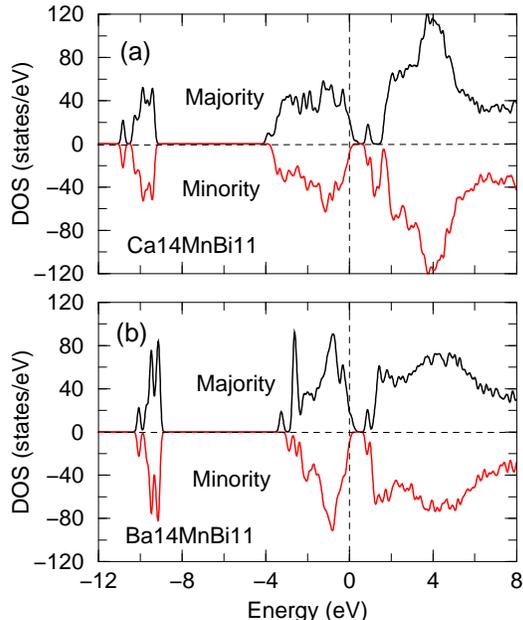}}
\caption{Total densities of states per unit cell for (a) 
ferromagnetically aligned
\ca~and, (b) \ba. The energies are referred to the Fermi level.
\label{TotDOS_FM}}
\end{figure}

The total density of states (DOS) of FM \ca~ and \ba~ are shown 
in {Fig.~\ref{TotDOS_FM}}~(a) and (b), respectively, which 
also reveal some general features. 
These plots have been obtained 
using three inequivalent k-points and a gaussian broadening of 0.1~eV.
Tests indicate 
that using more complete k-sampling (up to 12 inequivalent 
k-points) produces almost
identical results.
The occupied valence bands are
4-4.5 eV wide and nearly filled.  
A `gap' of  $\sim$~1~eV separates the valence complex
from a continuous spectrum 
of unoccupied states coming mainly from the Ca-Bi hybridization.
Within the gap, 0.5~eV above the top of the valence band, we find
four narrow (0.2 eV) bands.
In the minority spin DOS, near 
the top of the gap for \ca, and confused with the
onset of the conduction band in \ba, we can also find a peak 
that contains 
the unoccupied Mn 3$d$ bands. 

The band complexes below the gap consist of 182 bands of each
spin for AFM states, for which the spin directions are equivalent.
For  FM order
there are 192 majority and 172 minority bands, as 
the majority bands contain 
twenty more bands 
(for the five occupied 3$d$ electrons of the four Mn atoms)
than the minority bands. The total number of bands
in the valence complex is consequently 
the same for FM and AFM states, four electrons lacking for
these compounds 
to become semiconductors.
The 172 bands not associated with the $d$ states of Mn 
have mainly Bi character.
44 of them, located approximately 9~eV below 
the Fermi energy, are due 
the Bi 6$s$ states. The remaining
128, which form the uppermost valence band complex, 
can be classified as Bi 6$p$ bands, exhibiting some degree of 
hybridization with Mn and Ca/Ba states. 
From the 44 Bi atoms present in the unit cell we could
expect to have 132 Bi 6$p$ bands, however. 
The 
four Bi 6$p$ bands left out from the valence complex
form the, already mentioned, narrow peak in the middle of the gap. 
These
bands correspond to very localized 
states associated with the Bi$_3$ chains (see below).

The DOS
at the Fermi energy (E$_F$) 
is higher for the majority spin, (Fig.~\ref{TotDOS_FM})
and 
both compounds
are very close to half metallic, with the minority 
bands almost fully occupied
(as discussed later).
From the specific heat Siemens {\it et al.}\cite{heatcapacity}
obtained the Fermi level DOS N(E$_F$) =
1.7$\pm$0.7 states/eV-atom for \ba, and 1.6$\pm$0.1 states/eV-atom
for the similar compound Sr$_{14}$MnBi$_{11}$. 
These are quite large values,
comparable to the DOS of $d$ band metals.
However, Siemens {\it et al.}
pointed out that contributions
to the specific heat coming from the nuclear hyperfine 
splitting, not taken into account in their analysis,
could obscure their results, leading to large 
apparent values of the electronic DOS.
In fact, our estimations of the DOS at E$_F$ are much smaller:
N(E$_F$) = 0.3~states/eV-atom for both Ca and Ba compounds with FM order.
For AFM order, N(E$_F$) = 0.40-0.46~states/eV-atom 
depending on 
the specific AFM alignment.
These values can be compared with the average DOS in the valence band
complex;
leaving aside 
the Mn bands, we have 256 bands for both spin orientations
distributed over a range of $\sim$4~eV, leading to an average 
$<$N(E)$>$=0.6~states/eV-atom.

\subsection{Mn Character}
\label{MnCharacter}

\begin{figure}[tbp]
\epsfxsize=5.0cm\centerline{\epsffile{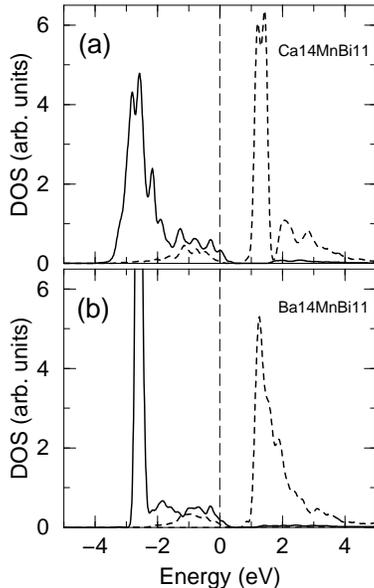}}
\caption{Densities of states projected into the 
Mn $d$ states in (a) \ca~and, (b) \ba.  The Fermi level defines
the zero of energy.  Solid lines denote majority; dashed lines denote
minority.  The exchange splitting is $\sim$4~eV in each case.
\label{Mn-d_FM}}
\end{figure}

Because of the magnetic moment on the Mn atom, we begin our discussion of
the electronic bonding with the Mn $d$ states. In all the cases we have
considered, the simple characterization is that the majority $d$ states are
filled and the minority states are empty, with an exchange splitting 
between them of $\sim$4 eV. Accordingly, the Mn atom can be 
described as
having a 3$d^5$ configuration or, alternatively, a 2+ valence.
The MnBi$_4$ tetrahedron is compressed along the $a-b$ plane and
only the $d_{xz}$ and $d_{yz}$ states are degenerate, so the crystal
field results in four
distinct $d$ levels. 
The crystal field splitting is negligible, however.  In 
\ca~(\ba),
shown in {Fig.~\ref{Mn-d_FM}}, the
majority $d$ states lie in a single peak centered 2.7 eV below 
E$_F$ with a width of
roughly 1 eV (respectively 0.4 eV).  The $d$ states do hybridize
with all states above this peak (with mainly Bi 6$p$ character)
to the gap just above E$_F$.  The minority
states are unoccupied and are concentrated in a peak 1.3 eV above E$_F$.
This unoccupied peak is narrow in \ca~(0.6 eV) and lies 
within the gap (see also Fig.~\ref{Bands_AFMCa}). In \ba~
it is much broader (at least 1~eV) meeting the bottom of the
conduction band, and exhibiting a higher degree of hybridization 
with the conduction states.

\subsection{Bi Character}

\begin{figure}[tbp]
\epsfxsize=7cm\centerline{\epsffile{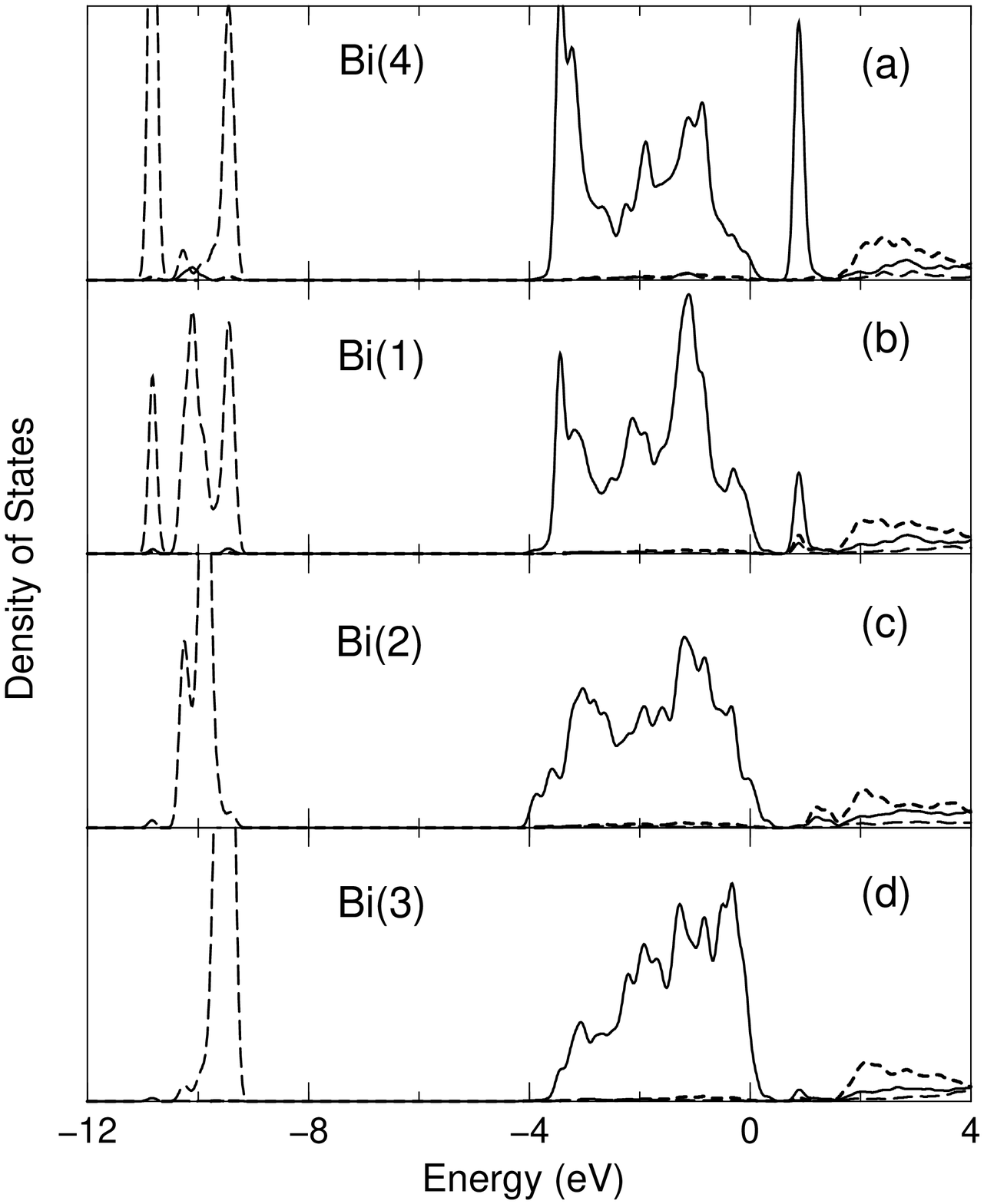}}
\caption{Densities of states projected into the 
$s$ (long dashed lines at lower energies), $p$ 
(solid line), and $d$ (dashed lines at higher energies)
symmetry orbitals of the Bi1, Bi2, Bi3, and Bi4 sites in
ferromagnetically aligned \ca.  The differences occur primarily in
the gap region 0-2 eV, and are discussed in the text.  Spin-averaged
spectra have been plotted; only the polarization on the Bi2 site
($\sim -0.06 \mu_B$) is appreciable.
\label{Bi1234_DOS}}
\end{figure}

The spin-averaged DOS projected into the
different types of Bi atoms 
in the unit cell can be found in 
{Fig.~\ref{Bi1234_DOS}}. 
The Bi 6$s$ bands lie 9-10~eV
below the Fermi level, and have negligible hybridization
with states of different character. 
The Bi 6$p$ states make up a large component of the valence bands,
exhibiting hybridization with both Mn and alkaline earth states,
and hence are crucial in the bonding and in the magnetic coupling.
The valence bands are 4 eV wide in \ba~and slightly (5\%) wider in
the smaller volume \ca~compound.  There is a `gap' of $\sim$1 eV between
the valence and conduction bands, {\it except} that a very narrow set of
unoccupied bands lies within this gap.
Although
we have neglected in {Fig.~\ref{Bi1234_DOS}}
the small effects due to the induced exchange splitting,
it should be kept in mind that the polarization
of the Bi 6$p$ states near the E$_F$
is crucial
to understand the magnitude and distribution
of the magnetic moment.

The complications are still considerable, since there are four distinct
Bi sites.  The Bi2 sites correspond to the 
distorted MnBi$_4$ tetrahedron, the Bi$_3$ linear chain
is formed by a central 
Bi4 atom bonded to two Bi1 atoms
and there are also four isolated
Bi3 atoms.  The first thing that is evident from the Bi spectra
in {Fig.~\ref{Bi1234_DOS}} is that there is not
a great deal of difference in the densities of states of the four Bi
sites, indicating that their charges 
are not very different.  The 
Bi1-Bi4-Bi1
linear unit has only a small (but nonzero) DOS at E$_F$, whereas both 
the Bi2 bonded to the Mn and the `isolated' Bi3 site have much
larger fractions of the Fermi level DOS. In the case of 
the Bi2 atoms
this DOS is almost entirely due to the majority spin contribution.

\begin{figure}[tbp]
\epsfxsize=6.0cm\centerline{\epsffile{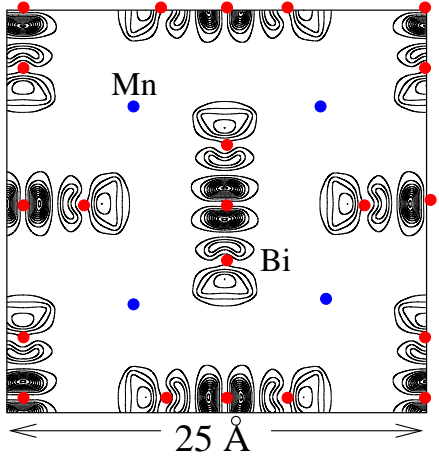}}
\caption[]{ Contour plot in the (001) plane 
of the density of the antibonding
$\sigma^{\star}$ states localized in
the Bi$_3$ chains of FM \ca. These
four bands are located within the
gap,
around 1~eV above the Fermi energy.
The contours begin at 3$\times$10$^{-4}$~e/Bohr$^{3}$
and increase by steps of 6$\times$10$^{-4}$~e/Bohr$^{3}$ 
(respectively, 7 and 14 electrons per unit cell).
The positions of the Mn and Bi atoms in the plane
are indicated schematically.
\label{antibonding}}
\end{figure}

The unoccupied `gap states', one state and therefore one
band for each Bi$_3$ unit (four in the unit cell), are 
associated with the Bi1
and Bi4 atoms only, with no appreciable spin polarization.
Since they lie in the gap and have a dispersion
of only 0.2 eV, these states are quite localized. 
{Fig.~\ref{antibonding}} shows
contour plots of their
density in a plane containing the Bi1-Bi4-Bi1 sites.
Their shape, 
with nodes in the bond region
between Bi1 and Bi4 atoms, and the 
fact that (see {Fig.~\ref{Bi1234_DOS}}) they have a larger
weight in the central Bi4 atom,
reveal them
as antibonding $\sigma$ molecular orbitals. The presence of 
these $\sigma^{\star}$ states 
and the negligible population of the
5$d$ states of the 
Bi4 atom, which rules out the possibility 
of a $dsp^3$ hybridization,
confirm the description of the Bi$_3$$^{7-}$ anion
as a hypervalent ``three center, four electron" bonded 
structure.
This type of bonding
was already predicted for As$_3$ in Ca$_{14}$GaAs$_{11}$ by 
Gallup {\it et al.}~\cite{gallup}, 
who applyed first-principles 
calculations to a simplified model (30-atoms unit cell)
of this compound.

The conduction bands contain the rest of the Bi 6$p$
spectral weight along with Bi 5$d$, Ca or Ba $s$, 
and Mn 3$s$, 3$p$ and minority 3$d$ contributions.

\begin{figure}[tbp]
\epsfxsize=8.0cm\centerline{\epsffile{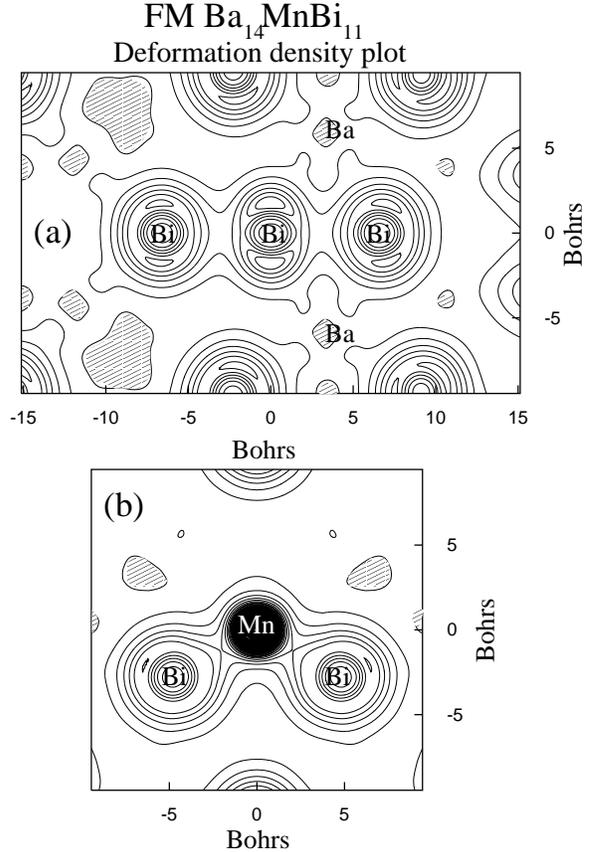}}
\caption{(a) Contour plot of the deformation density of ferromagnetically
aligned \ba~in a (110) plane, illustrating the Bi$_3$ unit bonding (top
panel).  The bottom panel (b) shows the (100)
plane through the MnBi$_4$ tetrahedron,
indicating the charge `deformation' is rather spherical around the atoms
and is therefore primarily charge transfer.  Only the region of 
increased charge is shown, with contours beginning at zero and increasing
by 0.03 $e$/\AA$^3$. The position of the atoms in the planes is
indicated by their chemical symbol. The shaded areas correspond to 
regions with defect of charge.}
\label{FM_Ba.def_dens}
\end{figure}

\subsection{Ca/Ba Character}
The formal valence picture suggests considering the Ca and Ba
atoms as dipositive ions.  This picture should not be taken too literally,
as we find some $s$, $p$, and $d$ alkaline earth contributions to the DOS
in the valence band region (figure not shown).  
The Ca/Ba $d$ DOS peaks lie 4 eV above $E_F$,
whereas the $s$ and $p$ contributions are spread over a large region 
from $\sim$-3 eV through and above 
the $d$ bands, with rather little structure
in their distribution.

\section{Charge Transfer and Formal Valence}
\label{formalvalence}

\begin{figure}[tbp]
\vskip 3mm
\caption{Greyscale plot in a (100) plane of
the (pseudo-)charge density associated to the unoccupied states
below the gap (holes) in AFM \ca. Panel (a) shows the
spin up holes, and (b) the spin down holes.
The arrows indicate schematically the positions of the
Mn atoms
and the orientations of their atomic magnetic moments.
The positions of one of the Bi2 and 
Bi4 atoms 
are also indicated.
We have used a logarithmic scale which
saturates to black at 10$^{-3}$~e/Bohr$^3$ and
to white at 10$^{-5.3}$~e/Bohr$^3$
(respectively, 22 and 0.1 electrons per unit cell)
\label{AFM_holes}}
\end{figure}

Now we address the question of whether the formal valence that
the Zintl picture associates to the different groups of atoms
provides a reasonable description of our results.
The formal valences are: Ca$^{2+}$/Ba$^{2+}$, 
isolated Bi$^{3-}$,
Bi$_3^{7-}$ linear chains, and MnBi$_4^{9-}$
tetrahedra.
These latter two are very large charges and,
as can be expected,  
the Mulliken populations are much smaller,
although still indicative of large charge
transfers. For example for 
\ca~we obtain: Ca$^{0.8+}$, MnBi$_4$$^{3.7-}$, Bi$_3$$^{2.8-}$, 
Bi$^{1.2-}$. The main reason for 
this difference between formal and Mulliken charges
can be traced back to the fact
that, while there is not a single state with pure or main Ca/Ba character,
most of the states in the valence band exhibit some small hybridization
with the orbitals of the alkaline earth 
atoms, specially those at the top
of the valence band with the  
$d$ states of Ca/Ba. 
In fact, the $s$ population of the
alkaline earth ions (the only orbital that would be included
in a simplified description of these atoms)
is only 0.4 electrons
for both Ca and Ba compounds, picturing them as
Ca(Ba)$^{1.6+}$.
Futhermore, Ca/Ba components 
are generally very small
for each individual eigenstate of the solid, and
formally they can be safely considered as dipositive.

{Fig.~\ref{FM_Ba.def_dens}} shows the contour plots 
of the deformation density (difference between 
the self-consistent charge density and
that obtained from the 
sum of spherical neutral atom contributions)
in different planes for FM \ba. Two things can 
observed: {\it i)} there is a great deal of charge
transfer to the Bi atoms 
from the Ca/Ba atoms, which are always 
surrounded by a region showing a depletion of charge; 
{\it ii)} 
the deformation density is quite featureless and spherical 
around the 
Bi atoms, indicating that indeed the ionic charge transfer is the
main mechanism involved. 

We now consider the polyatomic 
anions. The situation of
Bi$_3$ is specially clear.
Two antibonding molecular states 
(one per spin) remain
unocuppied and 
the valence states are almost completely occupied up to the gap,
with a very small contribution to the 
Fermi level DOS (see {Figs.~\ref{Bi1234_DOS} 
and \ref{antibonding}}). All these
data 
together identify 
the Bi$_3$ unit as having a 7- formal valence.
The weight of the gap peaks in
{Fig.~\ref{Bi1234_DOS}} suggest 
that less 
charge is transfered to the central Bi4 
than to the Bi1 sites, which  
is confirmed
by the Mulliken analysis (around 0.3~$|e|$ less). 
This completely
agrees with the ``three center, four electrons" 
model of bonding proposed for the Bi$_3$ 
unit~\cite{gallup}.

We now discuss the cases of the MnBi$_4$ units and the isolated Bi3 ions.
On one hand, a 3- formal charge 
would correspond to closed shell Bi3 anions, 
with a negligible contribution to the DOS at E$_F$.
On the other hand, considering the 2+ effective valence 
($d^5$) of the Mn atom, 
the MnBi$_4$ group would be formally able to accept up to ten electrons.
This implies that a MnBi$_4^{9-}$ tetrahedron would 
be lacking one electron to be closed shell.
This  
agrees with our analysis of the electronic structure (Section~\ref{TotDOS}),
which indicates
that these compounds lack four electron per unit cell (i.e. one electron
per MnBi$_4$ group) to become semiconductors.
Furthermore, taking into account the calculated 
reverse polarization of the Bi2 sites, 
the hole associated to each MnBi$_4^{9-}$ 
tetrahedron should be expected to be 
parallely aligned with the atomic moment of the corresponding
Mn. This is confirmed by the data presented in {Fig.~\ref{AFM_holes}}, 
where a plot of the density associated to the 
holes for AFM \ca~(alignment type (a) of those shown in
{Fig.~\ref{Fig2}})
in the (100) plane is shown.  This plane contains Mn atoms,
some of the Bi2 atoms bonded to them, 
and Bi4 sites.
We can verify how
the holes, while not exactly confined to,
are fairly localized in the tetrahedra: the spin up holes
in those tetrahedra where the atomic moment of Mn points
in the up direction, and conversely for spin down holes.
An obvious consequence of  
this polarization of the valence 
band holes is the (close to) half metallic character of the 
band structure of the FM compounds, which will be 
examined in detail in the next section.
Hence we can conclude that the indentification of 
a valence 9- for the MnBi$_4$ unit and a valence 3- for the Bi3 ions 
are meaningful, and 
from the point of view
of a general picture our results
are consistent with counting of formal charges.

\section{Electronic states near the Fermi Level}

\subsection{Character of the states}

\begin{figure}[tbp]
\epsfxsize=6.5cm\centerline{\epsffile{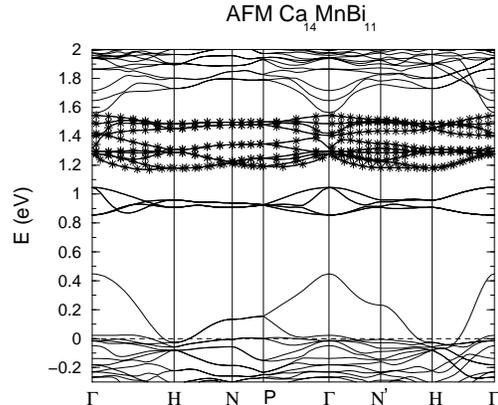}}
\caption{Band plots along primary symmetry
directions for antiferromagnetically
aligned \ca.  The narrow bands at 0.8-1.1 eV  arise from
the antibonding $\sigma^{\star}$ states
localized in the Bi$_3$ chains. Those marked with
stars are formed by the unoccupied $d$ states of Mn.
The energies are referred to the Fermi level.
\label{Bands_AFMCa}}
\end{figure}

For simplicity we will first discuss the case of the
AFM order.
{Fig.\ref{Bands_AFMCa}} shows the
band structure for AFM \ca~ (alignment type (b) of those shown in
{Fig.~\ref{Fig2}}).
There are two bands unfilled below the gap, and hence
there are four electrons
less (i.e. one per formula unit)
than required to make this compound a semiconductor. 
The figure shows with detail
the region of the
`gap' between bonding
and extended
conduction states. For each spin orientation this gap
contains two groups of bands corresponding
to very localized states:
one very narrow group of four bands
coming from the antibonding
states of the Bi$_3$ chains in the energy range 0.85-1.05~eV,
and ten unoccupied 3$d$ Mn
bands in the range 1.2-1.6~eV, the last ones
almost degenerate with the begining
of the continuum of conduction states.
The energy position of these groups of bands is very similar for
AFM and FM states, but while the Bi$_3$ bands do not show any spin
polarization, of course for FM order the twenty
unoccupied Mn 3$d$ bands only occur
in the minority band structure.

\subsection{Half metallicity of FM compounds}

\begin{figure}[tbp]
\epsfxsize=8.0cm\centerline{\epsffile{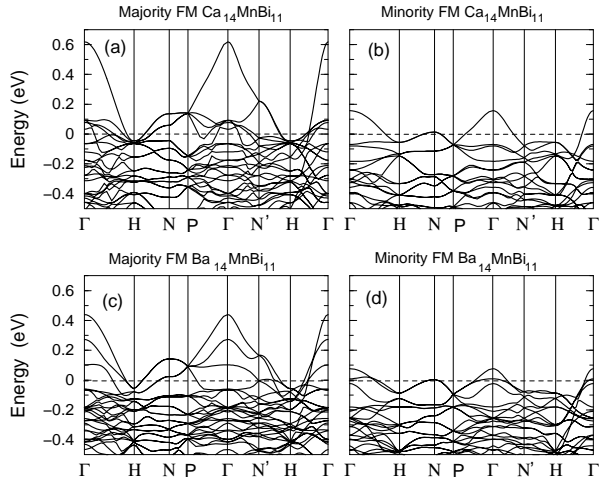}}
\caption{Band plots in the vicinity of the Fermi level
along primary symmetry directions for ferromagnetically
aligned \ba~and \ca.  The symmetry points correspond to the
tetragonal BCC unit cell. The energies are referred to the
Fermi level.
\label{Bands_FMCaBa_Ef}}
\end{figure}

The bands for both FM \ca~and \ba~
near E$_F$, along several symmetry
directions, are shown in {Fig.~\ref{Bands_FMCaBa_Ef}}.
Except for details
(which affect Fermi surface shape and therefore might become important)
the bands are similar and we discuss their common characteristics.
The band structure is very close to half metallicity.
The minority bands are occupied except for a $\Gamma$-centered pocket
that contains only a fraction of a hole (around 50\%).
The rest of the four holes correspond to the majority
bands, which have several partially occupied bands:
five in \ca, four in
\ba.

The almost half metallic band structure is
especially interesting in light of the reported colossal
magnetoresistance in the related compound
Eu$_{13.97}$Gd$_{0.03}$MnSb$_{11}$.~\cite{chan}
The CMR
system La$_{1-x}$D$_x$MnO$_3$, with D = Ca, Sr, or Ba, and $x\sim \frac
{1}{3}$, are believed to be half metallic or close to it~\cite{book}.
Eu$_{13.97}$Gd$_{0.03}$MnSb$_{11}$ represents a case in which a
compound, Eu$_{14}$MnSb$_{11}$, that is isovalent with the ones studied here
is doped with additional carriers (Eu$^{2+} \rightarrow$ Gd$^{3+}$), which
will drive it toward half metallicity.

\subsection{Magnetic Couplings}

\begin{figure}[tbp]
\vskip 3mm
\caption{Greyscale plot in a (100) plane of 
the (pseudo-)charge density near E$_F$ in FM \ca. Top panel (a)
for occupied states, and bottom panel (b) for holes. 
The positions of one of the Bi2, Bi4 and Mn atoms 
in the plane
are schematically indicated.
Note the
bonding chains coupling Mn atoms (darkest spots) formed 
by the two intervening Bi2
atoms. We have used a logarithmic scale which
saturates to black at 10$^{-2.8}$~e/Bohr$^3$ and
to white at 10$^{-5.3}$~e/Bohr$^3$ 
(respectively, 35 and 0.1 electrons per unit cell)
\label{near_Ef}}
\end{figure}

An RKKY model for the magnetic interaction is the simplest assumption to
make in a dilute magnetic metal, and in the 
absence of other information, that
is what has been used so far in the interpretation of magnetic coupling
in these 14-1-11 magnets.~\cite{reviewZintl}
Our identification of 
near neighbor couplings ({Table~\ref{table1}}) invalidates
a spherical RKKY model, where the coupling strength depends only on
distance.  This conclusion is supported by {Fig.~\ref{near_Ef}}, which
shows a grayscale plot of the (pseudo-)charge density from states 
near E$_F$ in
\ca~in a (100) plane containing -Mn-Bi2-Bi2-Mn- bonding chains. 
In this figure, panel (a) contains the density from states in the range
from E$_F$-0.5 eV to E$_F$.  Panel (b) contains the charge density
from the unoccupied states between E$_F$ and the gap (holes).  
Both panels clearly show that the density associated to 
the states near E$_F$ concentrates along the -Mn-Bi2-Bi2-Mn- 
chains, making them the most likely conduit for 
the exchange coupling between Mn atoms.
The occupied
states, panel (a), shows direct charge peaks along this bonding chain,
although the states may be predominantly antibonding.  Panel (b) also
indicates density along this chain, and is apparently more antibonding,
since the charge maxima centered on the Bi2 sites do not point toward 
the neighboring Mn atom. However, this has to be taken with some
caution 
since, as already discussed in {Section~\ref{structure}}, 
other atoms
may intervene in the bonding
between neighboring MnBi$_4$ units. In particular, 
three Ca/Ba 
common nearest neighbors of those Bi2
atoms in neighboring connected tetrahedra, which are
not contained in the (100) plane shown.
In fact, the charge maxima 
in {Fig.~\ref{near_Ef}} (b) 
seem to point from the Bi2 atoms 
to some of those 
alkaline earth ions located $\sim\pm$0.7~\AA\ over the plane.

The general picture of the magnetic couplings that arises from 
our calculations is the following:
Mn atoms are divided between two different
networks (defined by the -Mn-Bi2-Bi2-Mn- bonding chains);
the nearest neighbors 
coupling within the same network is 
the strongest and 
always FM; the couplings among Mn atoms belonging to
different networks are much smaller and 
may be AFM, explaining the 
experimental 
AFM ground state of \ba. In particular, 
the presence of two internally 
FM coupled networks is consistent with the
observation that, although it suffers an AFM transition, 
the Curie temperature obtained for \ba~from
the fitting of the high temperature magnetic 
suceptibiliy is positive.~\cite{kuro}
Therefore, although we have failed to predict
the antiferromagnetism of \ba, we expect that
the magnetic alignment in this compound must be
quite similar 
to the most stable of the AFM orders discussed in
this paper ({Fig.~\ref{Fig2}} (b)),
i.e. one network entirely spin up, and the
other entirely spin down.

\section{Discussion: Study of 
Charged MnBi$_4$ Clusters}

The results of the previous sections
point out that, rather than the Mn ions,
the MnBi$_4$ tetrahedra should be considered as the {\it magnetic units}
in these compounds. The amount of charge transferred 
to the MnBi$_4$, in combination with the effective 
valence of the Mn atom, also
plays a crucial role to understand the magnetic moment, the conduction
properties, and the interplay between them. For this reason we 
devote a separated section to the study
electronic
structure of free standing
charged MnBi$_4$ tetrahedral clusters and how these results 
translate
into
implications for the solid. 

\begin{figure}[tbp]
\epsfxsize=8.0cm\centerline{\epsffile{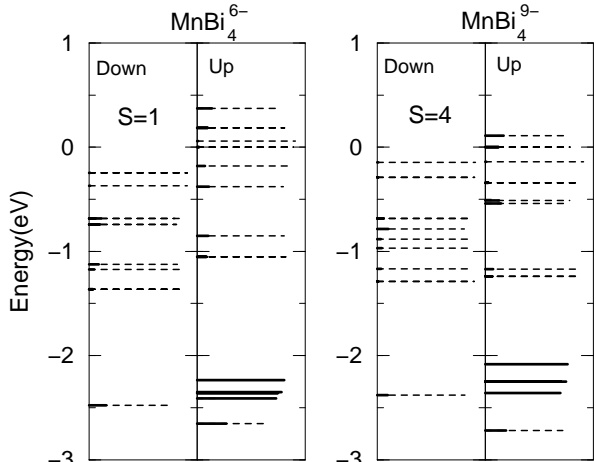}}
\caption[]{ Energies of the molecular orbitals of the
MnBi$_4^{9-}$ and
MnBi$_4^{6-}$ tetrahedral
clusters. The zero of energy is located at
the highest occupied molecular orbital.
The cluster geometry was fixed to that
of the MnBi$_4$ tetrahedron in crystalline \ba.
See text for notation.}
\label{MOenergies}
\end{figure}

{Fig.~\ref{MOenergies}} shows some of
the energy levels of
clusters with a total net charge of -9~$|e|$ (nominal for the
MnBi$_4$ group in the Ae$_{14}$MnBi$_{11}$ compounds), and
a smaller charge of -6~$|e|$.
The geometry of the cluster was kept fixed
to that of a MnBi$_4$ tetrahedron in crystalline \ba.
Due to the use of periodic boundary conditions,
it was necessary to assume
a uniform neutralizing background in the
calculations.
The horizontal scale represents the
projection of each molecular orbital
(MO) into the different atomic species,
solid
and thinner dashed lines
correspond to Mn and Bi$_4$ contributions, respectively
(which sum to unity). Different degeneracies
of the levels have not been taken into account
in making the figure. 
Only the group of levels close to the
HOMO is shown, corresponding to MOs
mainly due to the Mn 3$d$
and Bi 6$p$ states. Therefore,
in the range of energies shown in {Fig.~\ref{MOenergies}},
there are twelve MOs for both spin orientations, and
five additional MOs for the majority spin (only three levels 
are seen in Fig.~\ref{MOenergies} due to the degeneracies) 
with main character
in the Mn 3$d$ states, which corroborates the $d^5$ configuration
of the Mn atom also in the case of the free standing 
MnBi$_4$ clusters. Only one of the Bi 6$p$ MOs has a larger
binding energy than the Mn 3$d$ states. This MO has
a clear bonding configuration, with
a $p$ orbital from each Bi atom pointing in phase
towards the central Mn, and includes some hybridization
with
the $s$ symmetry orbitals of Mn.
Above the energy levels shown in {Fig.~\ref{MOenergies}}
there is a `gap' of $\sim$2~eV ($\sim$3~eV)
for the minority (majority) spin, and
well below the HOMO ($\sim-9$~eV)
four MOs coming from the Bi 6$s$ states can be found.

The most striking fact of the electronic structure presented
in {Fig.~\ref{MOenergies}} is the
{\it half metallic} character it suggests: the minority spin valence
states are all occupied for both clusters, lying at least
0.2~eV below the HOMO.
This situation holds for a large range
of net charging of the MnBi$_4$ clusters,
showing the robustness of the half metallicity.
The most probable origin of this
polarization of the Bi states is
the Pauli exclusion principle:
the majority spin MOs have to be orthogonal to the
localized Mn 3$d$ orbitals and the corresponding increase
in
kinetic energy pushes these levels
to higher energies than those of the minority
spin states.  As a result, the Bi $6p$ states are polarized
oppositely to the Mn $3d$ moment.

\begin{table}[tbp]
\caption[]{ Total magnetic moment S ($\mu_B$), as a functions
of the net charge Q$_e$ ($|e|$) and total number of
valence electrons N$_e$, for MnPbBi$_3$,
MnBi$_4$ and, MnBi$_3$Po tetrahedral clusters
with a central Mn atom. The cluster geometry
was kept to that
of the MnBi$_4$ tetrahedra in crystalline \ba.}
\begin{tabular} {c|cc|cc|cc}
       & \multicolumn{2}{c|}{MnPbBi$_3^{Q_e-}$} &
       \multicolumn{2}{c|}{MnBi$_4^{Q_e-}$ } &
       \multicolumn{2}{c}{MnBi$_3$Po$^{Q_e-}$ } \\
Q$_e$   & N$_e$ & S & N$_e$ & S & N$_e$ & S \\
\hline
-6 &  32  & 0 & 33    & 1 & 34    & 2  \\
-7 &  33  & 1 & 34    & 2 & 35    & 3   \\
-8 &  34  & 2 & 35    & 3 & 36    & 4  \\
-9 &  35  & 3 & 36    & 4 & 37    & 5  \\
-10 & 36  & 4 & 37    & 5 & --    & -- \\
-11 & 37  & 5 & --    & -- & --   & -- \\
\end{tabular}
\label{clusterSpin}
\end{table}

As a consequence of the half metallic electronic structure,
the total spin
of the
tetrahedral MnBi$_4$ clusters
depends on the total
number of valence electrons, as shown in
{Table~\ref{clusterSpin}}.
Taking into account the $d^5$ configuration 
obtained for the
Mn atom,
the MOs
due to the 6$p$ Bi
states can still accomodate up to ten extra-electrons.
In that case, with a net charge of -10~$|e|$,
both majority and minority states are fully occupied
and the magnetic moment of the cluster equals that of the
Mn 3$d$ shell (5$\mu_B$).
If the net charge of the cluster is then reduced, the holes are always
created in the majority spin MOs and, as a consequence, the total
spin of the cluster is reduced.
In particular,  the formal charge
of the MnBi$_4$
group in the manganites studied in this paper is -9~$|e|$
(see section~\ref{formalvalence}),
which leaves one unocuppied majority-spin state.
Therefore, consistent with the electronic structure of the 
clusters and the formal valence,
we can expect a magnetic moment of $\sim$4~$\mu_B$
on each MnBi$_4^{9-}$ tetrahedron, as has been
obtained in our calculations for the solids.
The data in {Table~\ref{clusterSpin}} also show that,
if the charge transfer to the Mn tetrahedra is lower than
9 electrons, even a larger reduction of the magnetic
moment could be expected.
The robustness of the dependence of the total spin
on the number of electrons is also shown
by substituting one of the Bi atoms by Pb, which
provides one electron less, and by a Po atom, which
provides one electron more than Bi to the cluster
valence population.

Since the rest of the atomic groups in the
unit cell are formally closed shell, we can expect the
main properties of the crystal
to arise from
MnBi$_4$ clusters, at least as far as
the interaction between MnBi$_4^{9-}$ units is weak
enough in the 
solid~\cite{hopping}. With such simplified model 
we can already anticipate the following properties:

{\it i}) the presence of a polarized hole  
localized in each MnBi$_4$ tetrahedron;

{\it ii}) as a consequence of the reverse 
polarization of the valence band
holes, the magnetic moment should be
reduced considerably from the Mn$^{2+}$
value to near 4~$\mu_B$ and,

{\it iii}) the FM order should exhibit a half-metallic band structure
with approximately four unoccupied majority spin bands (one for
each Mn atom in the unit cell);

{\it iv)} the hopping of these holes between neighboring 
tetrahedra should be the main mechanism of electric transport. 

These results are in fact corroborated by the calculations for the solids
presented in the previous sections, that take
into account the full interconnectedness of the lattice.

\begin{table}[tbp]
\caption[]{ Population of the Mn $d$ shell and magnetic
moments of different groups of atoms, as obtained
from a Mulliken population analysis for FM \ba,
and the compounds obtained by substituting a
Bi atom in one of the MnBi$_4$ tetrahedra by a
Pb and a Po atom, respectively. In all the cases
the structure of the \ba~was maintained. The
data
correspond to the subtituted (MnBi$_3$X) groups,
while those in brakets correspond
the remaining MnBi$_4$ tetrahedra.
}
\begin{tabular}{l|ccc}
& \multicolumn{3}{c}{Ba$_{56}$Mn$_4$Bi$_{43}$X}   \\
& X=Pb & X=Bi & X=Po  \\
\tableline
Q$_d$   &  5.14 (5.15) &   5.15 &  5.16 (5.15)   \\
$\mu_d$          &  4.24 (4.33) &  4.32 & 4.34 (4.32) \\
$\mu_{Mn}$       &  4.56 (4.63)&  4.62 &  4.66 (4.62)\\
$ \mu_{MnBi_3X}$  &  3.73 (4.39) & 4.35 & 4.52  (4.32)  \\
$ \mu_{unitcell}$&  16.98  & 17.59  &   17.04
\end{tabular}
\label{population2}
\end{table}

In addition,
we have performed calculations for some fictitious
materials
that provide
further support to the idea that an argumentation
based on the MOs of the isolated clusters is still
meaningful for the crystal.
In one case
we have substituted a Bi atom by Pb in
one of the MnBi$_4$ groups
of the unit cell of \ba.
According
to {Table~\ref{clusterSpin}}
the magnetic moment of the resulting MnPbBi$_3^{9-}$
tetrahedron should decrease to 3~$\mu_B$ from
the 4~$\mu_B$ of the pure Bi cluster.
In {Table~\ref{population2}} we can see that,
in the solid,
this substitution is in fact accompanied by
a decrease of both the total
(by 0.61~$\mu_B$) and the local magnetic
moment, which becomes 3.73~$\mu_B$ for the
tetrahedron where the substitution has
been performed while is almost unchanged
for the rest of the tetrahedra (4.39~$\mu_B$).
The case of the Po substitution is a little
bit more complicated. The
magnetic moment of
MnBi$_3$Po$^{9-}$
should increase to
5~$\mu_B$. This is again reflected in the
results for the solid but
in this case the observed change is
more modest, 4.52~$\mu_B$ versus
4.32~$\mu_B$ (see {Table~\ref{population2}}).
This different behavior is related to two different
effects. In the first place,
the Po substitution adds one electron to the
tetrahedron which is probably accompanied
by a
delocalization of the associated MOs (Wannier functions
in the solid), thus the picture solely
based on the electronic structure of the isolated clusters
becomes less accurate.
Another effect comes from the fact that the band structure
of \ba, while very close to, is not totally half metallic.
With the addition of one electron the half metallicity
is reached, forcing the total magnetic moment to
have integer values, which consequently diminishes it
from
17.59 to 17~$\mu_B$ with the Po substitution.
As expected,
all the changes
of the magnetic moment observed
upon chemical substitution
occur without much modification of
either the charge or the moment associated to
the Mn atom, or its $d$ shell.

\section{Conclusions}

In this paper we have presented and analyzed
the results of first-principles, full-unit-cell, LSDA
calculations of
\ca~and \ba.  
The general agreement between our results and the 
available experimental information supports a microscopic
framework within which to understand these complex materials,
which show the simultaneous presence of localized 
Mn magnetic moments and metallicity.
We also provide evidence for new properties,
not yet corroborated 
by measurements, such as near half-metallicity
of the FM 
manganites.

We summarize our main conclusions:

{\it (i)} 
Electron counting based 
on the Zintl picture, which 
visualizes these materials as an ionically bound collection of 
dipositive alkaline
earth ions and negatively charged ions and covalently bonded 
polyatomic anions 
(MnBi$_4$$^{9-}$, B$_3$$^{7-}$, and Bi$^{3-}$),
is a good starting point for the description
of the electronic structure of these compounds,
as suggested
by Gallup {\it et al.}~\cite{gallup} 

{\it (ii)} The Mn ions should be regarded as divalent Mn$^{2+}$, rather than
the original picture that considers a $d^4$ configuration 
(3+ valence) for the Mn ion.
Charge balance is maintained by the introduction of one hole into the Bi $6p$
states in the MnBi$_4$ tetrahedron; a heuristic connection
Ga$^{3+}$(Bi$_4$)$^{12-} \leftarrow \rightarrow$ Mn$^{2+}$(Bi$_4$)$^{11-}$
can be drawn, which also relates the charge states in the tetrahedron to the
insulating or conducting character of the solid.  
The hole in the Bi$_4$ tetrahedron can hop to neighboring
MnBi$_4$ units, accounting for the (hole) conduction in \ca~and \ba.

{\it (iii)} In spite of the majority Mn 3$d$ states being fully occupied,
the experimental magnetic moment of $\sim$4~$\mu_B$ 
per formula unit (i.e. per Mn atom)
is recovered in our calculations, since the hole in the Bi$_4$ tetrahedron
is parallel to the Mn moment (the unpaired electron is antiparallel).

{\it (iv)} The FM compounds are
close to half metallic, with the minority bands almost completely 
occupied. 

{\it (v)} The AeMPn crystal structure introduces its own peculiar magnetic
coupling. The MnBi$_4$ tetrahedra (the {\it magnetic units} 
of these compounds) are distributed in two disjoint interpenetrating
three dimensional networks. 
Magnetic coupling is  
ferromagnetic within each network, and AFM order arises 
if there are AFM interactions
between tetrahedra from different networks. These  
anistropic magnetic couplings
cannot be understood within a traditional, isotropic RKKY scheme.

{\it (vi)} Our calculations also provide some insight into 
the puzzle of how the
transition temperatures can be as high as 70~K for this family
of compounds when 
the Mn-Mn 
are more than 10~\AA\ apart. One point is that the Mn-Mn 
distance is not the meaningful parameter,
rather the relevant
parameter is the Bi2-Bi2 distance which determines the
hopping along the -Mn-Bi2-Bi2-Mn- chains. 

{\it (vii)} We speculate that full three dimensional ordering is enhanced 
by the comparatively strong FM coupling within each network, analogously
to the situation in layered cuprates such as La$_2$CuO$_4$ which order 
magnetically above room temperature in spite of very weak interlayer
exchange coupling\cite{La2CuO4}.

Our new picture of Mn in the AeMPn structure should impact the interpretation
of the valence fluctuation behavior of Yb$_{14}$ZnSb$_{11}$, where indications
are that both Yb$^{2+}$ and Yb$^{3+}$ ions are present\cite{prlFisher}.  
Dipositive Zn
destroys the electron count that would result in semiconducting behavior
({\it viz.} Ca$_{14}$Ga$^{3+}$As$_{11}$).  The system accommodates by
drawing an electron from the Yb ions, which is possible because of the
near degeneracy of the Yb$^{2+}$ and Yb$^{3+}$ configurations.  However,
it may also be possible to draw an electron from the $5p$ cluster orbitals
of some of the Sb$_4$ tetrahedra, as is done in every Bi$_4$ tetrahedron
in \ca~and \ba, with an accompanying unbalanced spin.  Further study will
be required to develop a consistent picture of this fluctuating valence
system.

Our description also gives new
insight into the interplay between the conduction and magnetic properties, 
and may play a crucial role in the CMR behavior observed 
in some of the 14-1-11 manganites.
In particular, it provides
a simple picture of the pressure induced ferromagnetic
transition observed in Sr$_{14}$MnAs$_{11}$~\cite{PRBpressure}, 
which seems to be accompanied
by a semicondutor-metal transition, and
was previously interpreted as signature of a RKKY type 
of magnetic interaction.
Compressing
the AeMPn structure causes an increase in the interation between the
MnBi$_4$ units, with the corresponding increase of the hopping probability
of the localized valence band holes. These holes 
are simultaneously responsibles of the
conduction process  
and vehicles of the 
exchange interaction between Mn 
atoms and, therefore, a decrease in the resistivity 
of these systems 
is necessarily accompanied by stronger 
magnetic interactions.

\section{Acknowledgments}
D.S.P acknowledges useful discussions with P. Ordej\'on.
D.S.P and R.M.M acknowledge support from the grants
No. DOE-8371494, and No. DEFG02/96/ER 45439.
D.S.P also acknowledge support from the Basque Government 
(Programa de Formaci\'on de Investigadores).
W. E. P. was supported by National Science Foundation Grant No.
DMR-9802076.



\begin{references}
\bibitem{nomenclature} Here we will follow
the nomenclature used in Ref.~\onlinecite{kuro} to denote the 
inequivalent sites of the different species.
\bibitem{polarization}
Polarization functions are constructed perturbatively from the
original orbitals, and have the same
confinement radius.
\bibitem{cutoffmeaning} These cutoffs indicate the fineness of the
real-space grid where the electronic density is projected in order
to calculate the exchange correlation, and 
part of the self-consistent
Hartree contributions to the total energy and Hamiltonian matrix.
The energy cutoff is the maximum energy of a
plane wave that can 
be represented using such a grid.
\bibitem{hopping} For example, 
from the $\sim$0.3~eV difference between the
lowest unoccupied molecular orbital (majority spin) 
and the highest occupied minority-spin 
molecular orbital,
we can estimate that the
effective hopping between MnBi$_4^{9-}$ tetrahedra 
must be smaller than $\sim$0.1~eV
in order for the half-metallicity to survive in the solid.
\end{references}
\end{document}